\documentclass[12pt,a4paper]{article}
\usepackage[utf8]{inputenc}
\usepackage{graphicx}
\usepackage{amsmath}
\usepackage{amssymb}
\usepackage{braket}
\usepackage{authblk}
\usepackage{mathrsfs}
\usepackage[dvipsnames]{xcolor}
\usepackage[miktex]{gnuplottex}
\usepackage{epstopdf}
\usepackage[toc,page]{appendix}
\usepackage[colorlinks,linktocpage]{hyperref}
\hypersetup{
	colorlinks=true,
	citecolor=blue,
	urlcolor=Plum
}
\title{Chaotic solitons in driven sine-Gordon model}
\author[1, 2]{D.~G.~Levkov\thanks{\texttt{levkov@ms2.inr.ac.ru}}}
\author[1, 2, 3]{V.~E.~Maslov\thanks{\texttt{vasilevgmaslov@ms2.inr.ac.ru}}} 
\author[1]{E.~Ya.~Nugaev\thanks{\texttt{emin@ms2.inr.ac.ru}}}

\affil[1]{\textit{Institute for Nuclear Research of the Russian Academy of Sciences, Moscow 117312, Russia}}
\affil[2]{\textit{Institute for Theoretical and Mathematical Physics, MSU, Moscow 119991, Russia}}
\affil[3]{\textit{Department of Particle Physics and Cosmology, Faculty of Physics, MSU, Moscow 119991, Russia}}
\date{July 8, 2020}

\begin{document}
	\begin{flushright}
		INR-TH-2020-018
	\end{flushright}
	\vspace{-1cm}
	{\let\newpage\relax\maketitle}
	\begin{abstract}
		Profiles of static solitons in one-dimensional scalar field theory satisfy the same equations as trajectories of a fictitious particle in multidimensional mechanics. We argue that the structure and properties of the solitons are essentially different if the respective mechanical motions are chaotic. This happens in multifield models and models with spatially dependent potential. We illustrate our findings using one-field sine-Gordon model in external Dirac comb potential. First, we show that the number of different ``chaotic'' solitons grows exponentially with their length, and the growth rate is related to the topological entropy of the mechanical system. Second, the field values of stable solitons form a fractal; we compute its box-counting dimension. Third, we demonstrate that the distribution of field values in the fractal is related to the metric entropy of the analogous mechanical system.
	\end{abstract}
	
	\textit{Keywords}: solitons; dynamical chaos; sine-Gordon model; fractals; metric (Kolmogorov-Sinai) entropy; topological entropy.
	\vspace{1cm}
	\section{Introduction and Summary}
	\label{IntroSum}
	There exists an amusing mathematical analogy between static
	solitons in one-dimensional field theory and point-particle trajectories in multidimensional
	mechanics. Indeed, the solitonic profiles typically
	satisfy second-order equations~\cite{rajaraman, rubakov, manton} 
	\begin{equation} 
	\label{eq:1}
	\frac{\partial^2 \varphi_i}{\partial x^2} = \frac{\partial	V}{\partial \varphi_i}\;,
	\end{equation}
	where $\varphi_i(x)$ are the fields of the model and $V(\varphi,
	\, x)$ is their scalar potential. These equations coincide with the Newton's law for the evolution in ``time'' $x$ of a fictitious particle with coordinates $\varphi_i(x)$ in an external potential $V_{mech} \equiv	-V$. Studying the mechanical trajectories, one can investigate the solitons.
	At $x\to \pm \infty$ the soliton fields approach the vacua~--- minima of the
	potential~$V$. Thus, the respective mechanical 
	trajectories $\varphi_i(x)$ lie on the separatrix: they start on the 
	maximum of  $V_{mech}(\varphi)$ at $x\to -\infty$ and climb onto the same or another 
	maximum in the infinite ``future''.
	
	In this paper we argue on the basis of the above analogy that one-dimensional
	static solitons have essentially different properties in models
	with multiple fields or models with position-dependent potential
	$V(\varphi,\, x)$ as compared to the simplest case of a single-field scalar theory. Indeed, mechanical
	motions are typically chaotic in models with several degrees of freedom. Smooth
	separatrix in this case is destroyed \cite{zaslavsky}, and the
	maxima of the potential $V_{mech}$ are connected by an infinite
	number of different trajectories. Since each trajectory represents the
	soliton, there exists an infinite number of the latter in the multifield
	models. Below we investigate such ``chaotic'' solitons and their distribution in the 
	configuration space. Notably, we find that many of these objects are linearly
	stable\footnote{Unlike the fictitious particle trajectories which 
		are unstable in the chaotic regime.} from the viewpoint of field theory: they cannot be destroyed by adding a small perturbation and time-evolving the resulting configuration. The subset of stable solitons is of our primary interest.
	
\begin{figure}[h]
		\centering
		\includegraphics{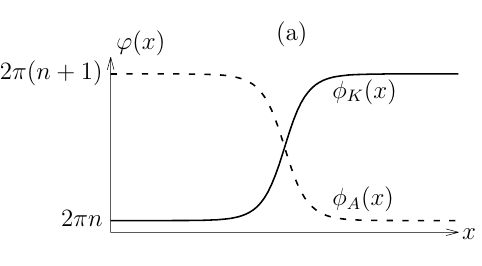}
		\includegraphics{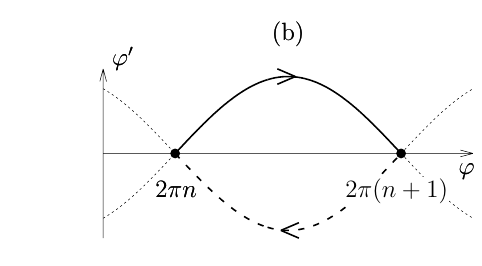}
		\vspace{-0.9cm}
		\caption{(a) Profiles of ``kink'' and ``antikink'' at constant $U$.
			(b) Respective trajectories in the mechanical ``phase space'' $(\varphi,\,
			\varphi')$. The ``time'' $x$ grows along arrows.}
		\label{fig:sin-Gordon}
	\end{figure}

	To be specific, we consider sine-Gordon model \cite{SineGordonBook} with coordinate-dependent potential~\mbox{\cite{malomed_jetp, malomed_physrevb}},
	\begin{equation}
	\label{osc+perturb}
	\varphi'' = \frac{\partial V}{\partial
		\varphi}\;, \qquad\qquad V(\varphi,\, x) = U(x) \, \left(1- \cos \varphi \right),
	\end{equation}
	where the prime represents $x$--derivative and $U(x)\geqslant 0$ is
	periodic\footnote{We study only static solitons 
		in this model, not their dynamics. The latter is also related to
		chaos, see~\cite{sinegordonsingledynamics, sinegordonsingledynamics2,
			sinegordonmultisolitondynamics, sinegordonperturbedkinkantikink, 
			sinegordonpertubedbreathers, sinegordonchaotic}.}. This model has vacua
	$\varphi_{n} = 2\pi n$, where $n$ is integer. If $U$ is a constant, the
	analogous mechanical motion is one-dimensional, conservative, and
	therefore integrable. In this case there exist
	only two types of static solitons: ``kink'' $\phi_K(x)$ and ``antikink''
	$\phi_A(x)$ interpolating between the
	neighbouring maxima of 
	$V_{mech} \equiv -V$, see Fig.~\ref{fig:sin-Gordon}a. The profiles of these
	objects form smooth separatrix (Fig.~\ref{fig:sin-Gordon}b) in the
	mechanical ``phase space'' $(\varphi, \varphi')$. Below we will consider
	nonintegrable case with spatially dependent $U(x)$. 
	
	It is worth noting that the sine-Gordon equation appears in several
	diverse setups. It describes relative phase difference between two
	coupled one-dimensional 
	superfluids at low energies \cite{condmatfirst, condmat,
		condmatnature}, rotation angle in classical ferromagnetic spin
	chain interacting with external magnetic field~\cite{mikeska, kumar, wycin, kawasaki}, or
	phase of superconductors in long Josephson
	junction~\cite{josephsondriven, malomed_jetp, malomed_physrevb}. In all these cases inhomogeneous
	potential can be achieved by spatial variation of parameters: external electric or magnetic fields, or impurities between the superconductors~\cite{josephsonwithdots}.
	
\begin{figure}[h]
		\begin{minipage}{0.5\linewidth}
			\includegraphics[scale=1]{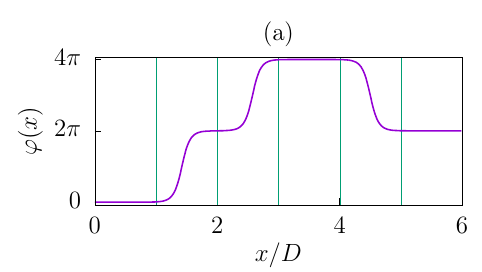}
		\end{minipage}
		\begin{minipage}{0.5\linewidth}
			\includegraphics[scale=1]{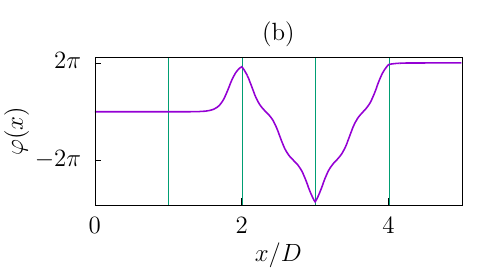}
		\end{minipage}
		\vspace{-0.2cm}
		\caption{Examples of static solitons (a) in the KAM
			regime at $\varepsilon=3 \times 10^{-7}$; (b) in the chaotic
			case at $\varepsilon = 3$. Vertical lines mark 
			positions of $\delta$-functions in
			Eq.~(\ref{perturb_as_delta}).} 
		\label{soliton_example}
	\end{figure}

	In numerical calculations we use the simplest dependence of
	the potential~(\ref{osc+perturb}), 
	\begin{equation}
	\label{perturb_as_delta}
	U(x) = 1 + \varepsilon \sum\limits_{m = -\infty}^{\infty} \delta(x - mD) \,,
	\end{equation}
	where $D=12$ is the period and the parameter $\varepsilon$ controls chaoticity of
	the underlying mechanical model. Although Eq.~(\ref{perturb_as_delta})
	may seem bizarre from the viewpoint of some applications, we expect 
	that our results remain qualitatively valid for any periodic modulation. For the potential~(\ref{perturb_as_delta}) the analogous mechanical motion is
	nearly integrable at $\varepsilon \lesssim 10^{-3}$. In this regime 
	Kolmogorov-Arnold-Moser (KAM) theory of  quasiperiodic
	motions~\cite{Arnold1, Arnold2, Moser} is 
	applicable, and the ``solitonic'' trajectories remain close to the separatrix in Fig.~\ref{fig:sin-Gordon}b. In fact, they toss erratically from vacuum to
	vacuum along this separatrix.
	The respective solitons can be
	obtained by matching together the kink and antikink profiles, see Fig.~\ref{soliton_example}a. The part of the ``phase space'' spanned by these
	trajectories, however, grows with $\varepsilon$ and fills a
	considerable region at $\varepsilon \gtrsim 0.1$. The solitons in the latter
	case appear in a wide variety of forms, see
	Fig.~\ref{soliton_example}b. 
	
\begin{figure}[h]
		\begin{minipage}{0.5\linewidth}
			\includegraphics[scale=1]{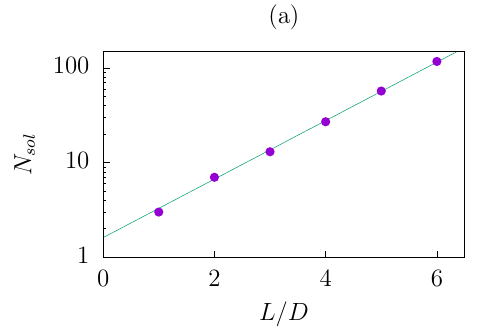}
		\end{minipage}	
		\begin{minipage}{0.5\linewidth}
			\includegraphics[scale=1]{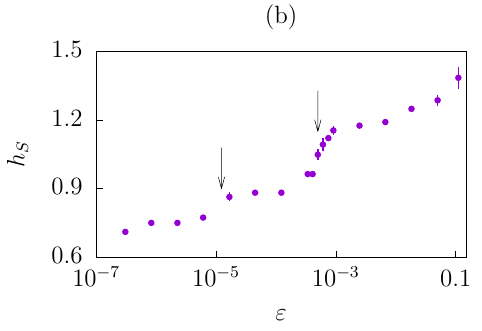}
		\end{minipage}
		\caption{(a) The number of stable solitons $N_{sol}(L)$ in a finite spatial
			box as a function of the box size $L$.  Numerical data (points)
			are fitted with Eq.~(\ref{eq:2}) (line). (b) The logarithmic
			growth rate $h_S(\varepsilon)$ as a function of the chaoticity
			parameter $\varepsilon$.} 
		\label{LnNsol(n)}
	\end{figure}

	In the main text we prove that the number of stable solitons $N_{sol}$ fitting in a
	finite spatial interval $0 \leqslant x \leqslant L$ grows exponentially with the interval
	size,
	\begin{equation}
	\label{eq:2}
	N_{sol} \propto \mathrm{e}^{h_S(\varepsilon) \, L/D}
	\qquad \text{as} \qquad  L \rightarrow +\infty, 
	\end{equation}
	where the growth rate $h_S(\varepsilon)$ monotonically increases with $\varepsilon$. The law~(\ref{eq:2}) is demonstrated numerically in
	Figs.~\ref{LnNsol(n)}a,~b. Steplike features of $h_S(\varepsilon)$
	(arrows in Fig.~\ref{LnNsol(n)}b) result from new types of solitons
	emerging at larger $\varepsilon$.
	
	Growth of the soliton multiplicity with $L$ can be easily explained in
	the KAM regime when the analogous mechanical motion proceeds along the
	smooth separatrix. In this case the stable solitons are completely specified by the
	set $\{\varphi_n\}$ of intermediate vacua. Say, the solitonic profile in
	Fig.~\ref{soliton_example}a corresponds to the sequence $\{\varphi_0,\,
	\varphi_1,\, \varphi_2,\, \varphi_2,\, \varphi_1\}$. The 
	number of possible sequences grows exponentially with their length
	$L/D$, and so does the number of stable solitons\footnote{At $\varepsilon \ll
		1$ some sequences do not correspond to soliton solutions. In the
		main text we account for these selection rules while deriving (\ref{eq:2}).}.
	
	In the main text we demonstrate that the growth rate $h_S(\varepsilon)$ of stable solitons is bounded from above by the topological entropy $h_T(\varepsilon)$ of
	the analogous mechanical system \cite{adlertopentropy, ottchaos},
	\begin{equation}
	\label{eq:3}
	h_S(\varepsilon) \leqslant h_T(\varepsilon)\;.
	\end{equation}
	The latter quantity characterizes complexity of the
	system i.e.\ diversity of its motions. 
	
\begin{figure}[h]
		\centerline{\includegraphics[scale=1]{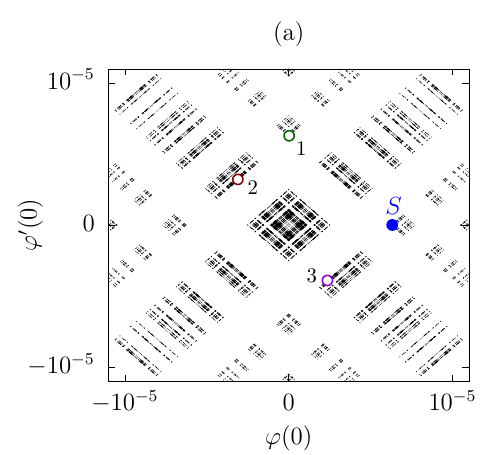}
			\includegraphics[scale=1]{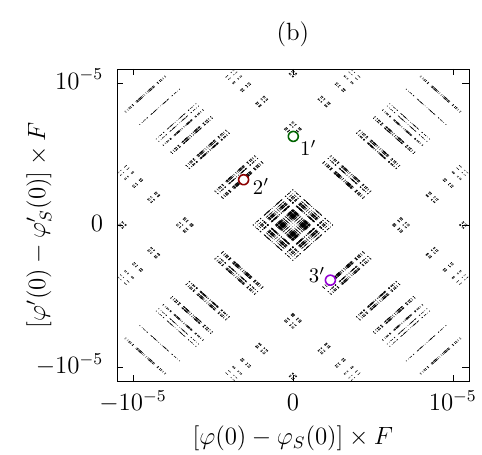}}
		\vspace{-0.2cm}
		\caption{(a) Field values $\varphi(0)$, $\varphi'(0)$ of 
			stable solitons at $x=+0$ in the model with
			$\varepsilon=3 \times 10^{-7}$. Only the region $|\varphi|, \, |\varphi'| <
			10^{-5}$ is shown. (b) Field values in the vicinity of the point S in
			Fig.~\ref{fractal2d}a magnified by a factor
			$F = \mathrm{e}^{\lambda_S(8D)}$, where $\lambda_S(8D) \approx 36.7$ is related to the
			Lyapunov exponent of the soliton~$S$.}
		\label{fractal2d}
	\end{figure}

\begin{figure}[h]
		\centering
		\includegraphics[scale=1]{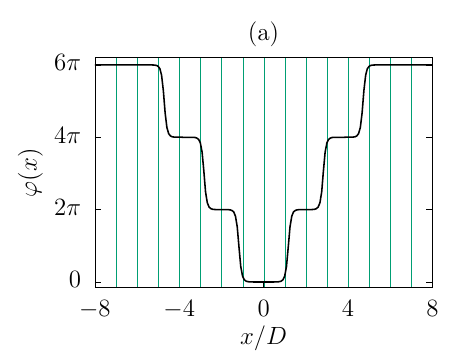}
		\includegraphics[scale=1]{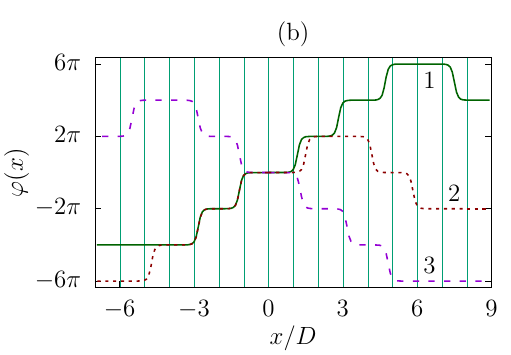}
		\includegraphics[scale=1]{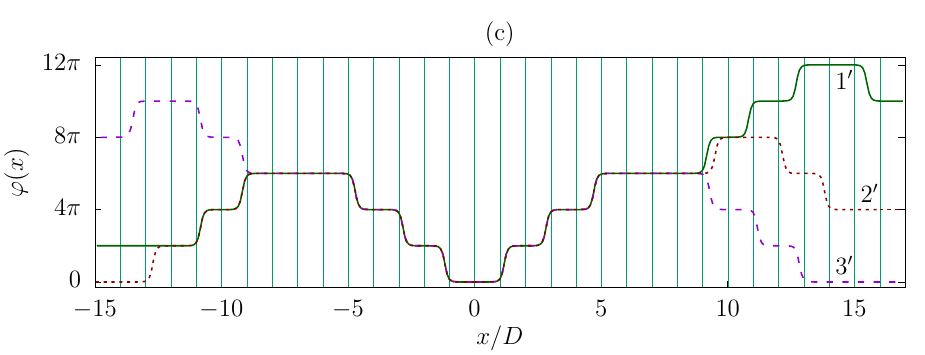}
		\vspace{-0.2cm}
		\caption{(a) The soliton with the Cauchy data S at $x=+0$, see
			Fig.~\ref{fractal2d}a.
			(b) The solitons 1---3 corresponding to empty circles in Fig.~\ref{fractal2d}a.  
			(c) The related solitons $1'$---$3'$ with Cauchy data
			in Fig.~\ref{fractal2d}b.}
		\label{selfsimilarsolst}
	\end{figure}

	It is well-known that distinct classes of trajectories are separated
	by fractal sets in the phase space of chaotic systems \cite{ottstrange, resonance0, fractal, Levkov:2007ce, Shnir}. We show that similarly, the solitons form a fractal in the space of 
	static field configurations\footnote{Recall that the soliton arriving
		to $\varphi_n$ at $x\to +\infty$ lies on the boundary between
		the solutions with $\varphi>\varphi_n$ and $\varphi<\varphi_n$ at large
		$x$.} $\varphi(x)$. To visualize the fractal, we compute the field
	values $\varphi(0)$, $\varphi'(0)$ of all stable solitons at a given
	spatial point $x=+0$ and plot them with dots in
	Fig.~\ref{fractal2d}a. For example, the point S represents the soliton
	in Fig.~\ref{selfsimilarsolst}a.
	
	We find that the set in Fig.~\ref{fractal2d}a is approximately
	self-similar. Indeed, it can be reproduced by magnifying a tiny region
	near one of its points, see Fig.~\ref{fractal2d}b. 
	To explain self-similarity, we choose points 1---3 in Fig.~\ref{fractal2d}a and related points $1'$---$3'$ in
	Fig.~\ref{fractal2d}b, then plot their profiles in
	Figs.~\ref{selfsimilarsolst}b and~\ref{selfsimilarsolst}c. Notably, at positive (or negative) $x$
	the solutions $1'$---$3'$ go along $S$
	first, then depart from it at $x \approx \pm 8D$ and follow the related profile 1, 2, or 3. Now, recall that the trajectory S is
	unstable from the mechanical viewpoint. Thus, small variations of its
	initial data $\varphi(0)$ and $\varphi'(0)$ lead to variations of the
	new ``initial data'' at $x=8D$ enhanced by a factor $\mathrm{e}^{\lambda_S(8D)}$, where
	$\lambda_S(x)$ is related to the Lyapunov exponent of $S$. One concludes that
	a small vicinity of every point in Fig.~\ref{fractal2d}a contains the entire set of
	``solitonic'' Cauchy data squeezed by the Lyapunov factor
	$\mathrm{e}^{-\lambda_S}$. 
	
	
\begin{figure}[h]
		\centering
		\includegraphics[scale=1]{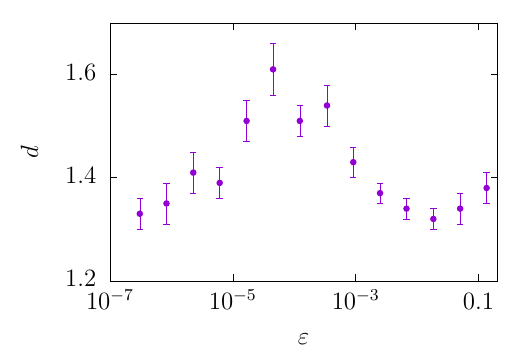}
		\vspace{-0.2cm}
		\caption{Box-counting dimension of the fractal in Fig.~\ref{fractal2d} at different $\varepsilon$.}
		\label{dbox(eps)}
	\end{figure}

	In Fig.~\ref{dbox(eps)} we plot the box--counting dimension~\cite{boxcountingorig, boxcountingreview}
	$d(\varepsilon)$ of the ``stable solitons'' fractal in Fig.~\ref{fractal2d}a at different values of the chaoticity parameter~$\varepsilon$. Apparently, $d$ is not integer\footnote{On the other hand, we will show that the field values $\varphi(0), \varphi'(0)$ of all solitons, both stable and unstable, form a dense set with fractal dimension $2$.}. Besides, it changes non-monotonically with $\varepsilon$ due to two competing effects. First, at larger $\varepsilon$ new solitons appear, increasing $d$. Second, Lyapunov exponents of already existing solitons grow with $\varepsilon$, making their field values closer in the $(\varphi, \varphi')$ plane. This effect decreases $d$ at large $\varepsilon$.
	
	In the main text we will demonstrate that at small $\varepsilon$ the fractal dimension $d(\varepsilon)$ is bounded from below by the stable solitons growth rate:
	$d(\varepsilon) \geqslant h_S(\varepsilon)/D$.
	
	An important characteristic of chaotic dynamics is the metric (Kolmogorov)
	entropy $K$. This quantity reflects divergence of the trajectories or,
	in other words, information growth rate during
	evolution \cite{ottchaos}. Positive values of $K$ indicate
	chaos. We suggest field-theoretic analogue $E$ of this quantity
	characterizing the distribution of stable soliton field values
	$\varphi(0)$, $\varphi'(0)$ at a given point $x=+0$. In particular,
	$E=0$ if all stable solitons have the same ${(\varphi, \,\varphi')}$. 
	Uniform distribution of solitonic field values gives $E=h_S$. In general case $E$ takes
	some value between these two limits, but it cannot exceed the
	Kolmogorov entropy of the underlying mechanical system, $E \leqslant
	K$. Thus, studying the soliton configurations one can investigate
	dynamical chaos in Eq.~(\ref{eq:1}).
	
	We perform explicit numerical computations only in the setup \eqref{osc+perturb}, \eqref{perturb_as_delta}. Nevertheless, we expect that the main qualitative properties of the solitons should be the same in other one-dimensional models with non-integrable static equations. Namely, the number of static solitons in these models should be infinite and the solitonic field values should form hierarchical structures in the configuration space. One can study the solitons using metric and topological entropies --- the instruments originally developed for dynamical systems.
	
	The paper is organized as follows. In Sec.~\ref{ModelSec}  we
	introduce the model (\ref{osc+perturb}) and discuss its 
	applications. The procedure of finding the solitons and determining their stability is described in Sec.~\ref{SecFindingSol}. Soliton multiplicity and its relation to the topological entropy are considered in Secs.~\ref{Multisoliton}
	and~\ref{SecTopEntropy}, respectively.
	The distribution of stable solitons in the
	configuration space is discussed in Sec.~\ref{SecFractal}.  The ``solitonic'' analogue of
	the metric entropy is suggested in
	Sec.~\ref{SecKolmEntropy}. Section~\ref{SecDiscuss}
	is devoted to conclusions and discussion of possible generalizations. 
	
	\section{The model: applications and properties}
	\label{ModelSec}
	We consider the theory of one-dimensional static scalar field with energy
	\begin{equation}
	\label{hamilt}
	H[\varphi] = \int dx \left( \frac{1}{2} (\partial_x \varphi)^2 + (1 - \cos \varphi) U(x)\right),
	\end{equation}
	where $U(x)$ is given by Eq.~(\ref{perturb_as_delta}). By definition, the solitonic profiles extremize this energy at its finite values, i.e. satisfy Eq.~\eqref{osc+perturb}. Stable solitons, in addition, correspond to local minima of $H[\varphi]$.
	They cannot be destroyed by adding a small perturbation and time-evolving the field in an energy-conserving way.
	
	Let us describe several situations where Eqs.~(\ref{hamilt}) and (\ref{osc+perturb}) appear. First, $(1+1)$-dimensional relativistic scalar field $\varphi(t, x)$ satisfies equation
	$\partial^2_t \varphi - \partial^2_x \varphi = -\partial V/\partial \varphi $
	that reduces to (\ref{osc+perturb}) in the static case for a particular choice of the potential $V$, see~\cite{rajaraman,rubakov}. The function $U(x)$ is then a time-independent external field.
	
\begin{figure}[h]
		\centering
		\hspace{-0.2cm}
		\includegraphics[scale=1]{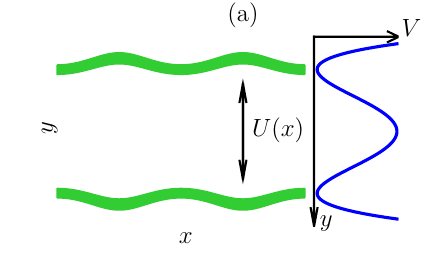}
		\hspace{0.5cm}
		\includegraphics[scale=1]{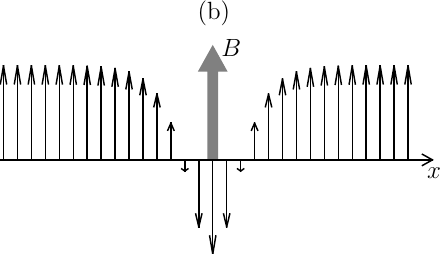}
		\vspace{-0.1cm}
		\caption{(a) Bose-Einstein condensate (thick solid lines) in a double-well potential with modulated coupling $U(x)$. (b) Ferromagnetic spin chain in a spatially inhomogenous magnetic field~$B$.}
		\label{condmat_examples}
	\end{figure}

	Second, one can consider Bose-Einstein condensate in the double well potential \cite{condmatfirst, condmat, condmatnature} forming two valleys stretched along the $x$ direction, see Fig.~\ref{condmat_examples}a. The condensates in the wells interact via tunneling through the potential barrier. It can be shown \cite{condmatfirst, condmat, condmatnature} that the relative phase difference $\varphi(x) = \arg \psi_1 - \arg \psi_2$ of the condensate wavefunctions $\psi_1, \psi_2$, satisfies (\ref{osc+perturb}), where $U$ characterizes coupling between the condensates.
	The spatial modulation can be introduced into this equation by periodically changing the barrier between the wells. Then the chaoticity parameter $\varepsilon$ in (\ref{osc+perturb}) is related to the modulation amplitude.
	
	The third example is a classical ferromagnetic spin chain arranged along the $x$ axis \cite{kumar, wycin} and interacting with the external magnetic field $B$, as depicted in Fig.~\ref{condmat_examples}b. Long-range dynamics of this chain can be described \cite{mikeska, kawasaki} by Eq.~(\ref{osc+perturb}), where $\varphi(x)$ is a rotation angle of spins in the $y-z$ plane. In this case $U \propto B/\left(J_s \Delta^2\right)$, 
	with $\Delta$ representing the interspin distance and $J_s$ characterizing interaction between the neighboring spins.
	Introducing spatial inhomogeneity into the magnetic field $B = B(x)$, we again obtain~(\ref{osc+perturb}).
	
	There are many other applications of the static sine-Gordon equation (e.g. \cite{josephsondriven, josephsonwithdots}), where spatial modulation of the potential can be achieved by variation of external parameters. Keeping these applications in mind, below we consider general properties of static solitons in the model (\ref{osc+perturb}),~(\ref{perturb_as_delta}) with\footnote{Note that $D$ is greater than the kink width in the pure sine-Gordon model, so the kink fits into a single period of $U(x)$.} $D=12$ and different $\varepsilon$.
	
	The trivial solutions to Eq.~(\ref{osc+perturb}) are the vacua $\varphi_n = 2 \pi n, n \in \mathbb{Z}$ corresponding to the absolute minima of energy (\ref{hamilt}) at all $\varepsilon$.\footnote{In particular, spatially inhomogeneous vacua do not exist.} In what follows we consider finite-energy solitons approaching these vacua at $x \rightarrow \pm \infty$.
	
\begin{figure}[h]
		\begin{minipage}{0.32\linewidth}
			\includegraphics{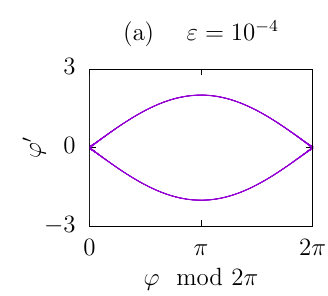}
		\end{minipage}
		\begin{minipage}{0.32\linewidth}
			\includegraphics{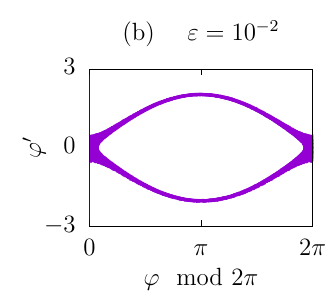}
		\end{minipage}
		\begin{minipage}{0.32\linewidth}
			\includegraphics{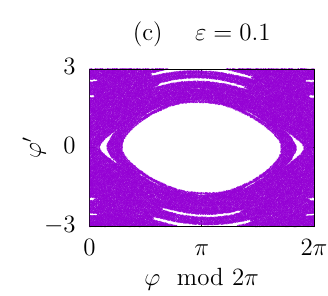}
		\end{minipage}
		\caption{Poincar\'e sections at different $\varepsilon$. The values of $\varphi$ are taken modulo $2\pi$.}
		\label{poincare}
	\end{figure}

	To illustrate chaos in Eq.~(\ref{osc+perturb}), we build the Poincar\'e sections~\cite{strogatz} for the analogous mechanical system at different $\varepsilon$. To this end we consider a generic solution $\varphi(x)$ starting from the vacuum $\varphi \rightarrow 0$ at $x \rightarrow -\infty$. Numerically evolving the solution, we plot the values of $\varphi, \varphi'$ at $x = mD+0$, i.e. after every period of the external potential, and obtain Fig.~\ref{poincare}. At $\varepsilon = 10^{-4}$ (Fig.~\ref{poincare}a) the solution remains close to the smooth curve --- the separatrix in Fig.~\ref{fig:sin-Gordon}b. In this case the analogous mechanical motion is nearly integrable. At $\varepsilon > 10^{-3}$ (Fig.~\ref{poincare}b) the values of $(\varphi, \varphi')$ already form a sizeable ``chaotic'' region near the destroyed separatrix. At even larger $\varepsilon$ in Fig.~\ref{poincare}c the solution $\varphi(x)$ tosses randomly inside the ``phase space'' covering a substantial part of it. Below we see how this chaos affects the soliton solutions in field theory.
	
	\section{Computing the solitons}
	\label{SecFindingSol}
	In this section we describe numerical method to find the solitons and determine their stability. 
	
	Since all vacua are equivalent, we consider only the solutions starting from $\varphi = 0$ at ${x \rightarrow -\infty}$. Near the vacuum Eq.~(\ref{osc+perturb}) becomes a linear Schr\"{o}dinger equation in the periodic potential $U(x)$~\cite{arnoldperiodic}. General solution of this equation includes exponentially growing and decreasing parts,
	\begin{equation}
	\label{general_hill_solution}
	\varphi \approx A \mathrm{e}^{\lambda_v x} f_A(x) + B \mathrm{e}^{-\lambda_v x} f_B(x)\,,
	\end{equation} 
	where $A, B$ are arbitrary constants, $\lambda_v = 1 + O(\varepsilon) > 0$ is the Lyapunov exponent of the vacuum, and $f_{A,B}(x)$ are periodic; we normalize them by $f_{A,B}(0) = 1$. We explicitly find $\lambda_v$ and $f_{A,B}$ in Appendix \ref{appendix_asymptotic}. Clearly, all solitons starting from the vacuum at $x \rightarrow -\infty$ have $B=0$, 
	\begin{equation}
	\label{left_condition}
	\varphi(x) \to A \, \mathrm{e}^{\lambda_v x} f_A(x) \quad \text{as} \quad x \to -\infty\,,
	\end{equation}
	and we parametrize them with the shooting parameter\footnote{In practice, $\ln A$ is more convenient at $A>0$; solitons with negative $A$ are then obtained by reflection $\varphi \rightarrow -\varphi$.} $A$.
	
	Analogously, the soliton profile arrives to some vacuum $\varphi_n$ at $x\to+\infty$, and its deviation from this vacuum is described by Eq.~(\ref{general_hill_solution}) with coefficients $A' \equiv 0$ and arbitrary $B'$. Taking the derivative of (\ref{general_hill_solution}), we obtain the boundary condition
	\begin{equation}
	\label{right_condition}
	\varphi'(x) = \left(-\lambda_v + \frac{f'_B(x)}{f_B(x)}\right) (\varphi - \varphi_n) \qquad \text{at} \qquad x \to +\infty.
	\end{equation}
	In what follows we solve Eq.~\eqref{osc+perturb} with boundary conditions \eqref{left_condition},~\eqref{right_condition}.
	
	We strongly rely on the shooting method. Imposing the boundary condition~(\ref{left_condition}) at $x = 0$, we numerically solve Eq.~(\ref{osc+perturb}) for every\footnote{To this end we change $A$ in small steps. We check that no solution is lost by changing the size of these steps.} $A$. Then we tune the value of $A$ to satisfy Eq.~(\ref{right_condition}) at large $x = L$. Once this is done, we have all solitons localized inside the interval\footnote{Expressions~\eqref{left_condition},~\eqref{right_condition} are more accurate if $\varphi(0)$ and $\varphi(L)$ are closer to the vacua. To increase precision, we perform computations on the larger interval $-D < x < L+D$, and then select solutions staying close to the vacua at $x<0$ and $x>L$.} $0 < x < L$.
	
	In the chaotic regime our solutions are exponentially sensitive to the initial data. Thus, we need an efficient and extraordinary precise numerical method to solve Eq.~(\ref{osc+perturb}).
	Our design of this method essentially relies on the simplified form of $U(x)$ in Eq.~(\ref{perturb_as_delta}). Namely, in the regions between the $\delta$-functions $mD < x < (m+1)D$ the potential $U$ is constant and Eq.~(\ref{osc+perturb}) can be solved explicitly in terms of elliptic functions, see Appendix \ref{appendix_solving}. At $x = mD$ one obtains matching conditions
	\begin{equation}
	\label{matching_sol}
	\varphi(mD+0) = \varphi(mD-0), \qquad \varphi'(mD+0) - \varphi'(mD-0) = \varepsilon \sin \varphi(mD).
	\end{equation}  
	As a result, our algorithm acts sequentially. Starting from $\varphi$ and $\varphi'$ at $x = mD+0$, it evolves them to $x = (m+1)D - 0$ using the explicit solution in Appendix \ref{appendix_solving}, then performs matching (\ref{matching_sol}) and proceeds to the next period of the potential.
	
	Importantly, we perform all calculations using arbitrary precision floating numbers with $40 \div 200$ digits. This gives us correct chaotic solutions of Eq.~(\ref{osc+perturb}) of arbitrary complexity.	
	The examples of these solutions are shown in Figs.~\ref{soliton_example} and \ref{selfsimilarsolst}. To study their statistical properties, below we obtain thousands of solitons of different forms and lengths.
	
	We are mainly interested in stable static solitons. They correspond to local minima of energy~(\ref{hamilt}). Adding small variation $\theta(x)$ to the solution $\varphi(x)$, one finds
	\begin{equation}
	H[\varphi + \theta] = H[\varphi] + \frac{1}{2} \int dx \, \theta(x) \hat{L}_\varphi(x) \theta(x) \,,
	\end{equation}
	where $\hat{L}_\varphi(x) = -\partial^2_x + \cos \varphi(x) U(x)$.
	Thus, the soliton $\varphi(x)$ is stable if the operator $\hat{L}_\varphi$ is positive-definite in the space of perturbations $\theta(x)$ vanishing at $x \to \pm \infty$.
	
	Numerically, we determine stability of solitons from the standard oscillation theorem. Namely, consider the perturbation
	\begin{equation}
	\label{Theta_0}
	\theta_0(x) = \frac{\partial \varphi(x)}{\partial A}\,,
	\end{equation} 
	where $\varphi(x)$ is a solution of (\ref{osc+perturb}) with the initial data (\ref{left_condition}). By construction, $\theta_0(x)$ satisfies $\hat{L}_\varphi \theta_{0} = 0$ and vanishes at $x \rightarrow -\infty$. Then by the oscillation theorem the number of its zeros equals the number of negative eigenvalues of the operator $\hat{L}_\varphi$.	
	In numerical code we compute $\theta_0(x)$ for each soliton and count the number of its roots. The soliton is stable if $\theta_0(x)$ is positive-definite. In Appendix \ref{appendix_stability} we explain how this calculation can be conveniently performed within our shooting procedure.
	\begin{figure}[h]
		\centering
		\includegraphics[scale=1]{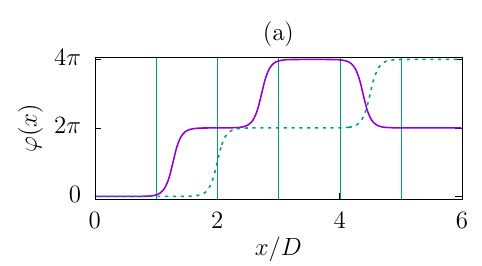}
		\includegraphics[scale=1]{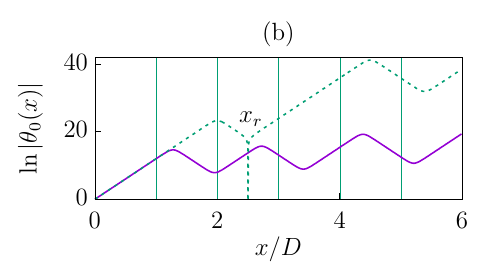}
		\vspace{-0.9cm}
		\caption{(a)~Stable soliton (solid line) and unstable soliton (dashed line) at ${\varepsilon = 4.5 \times 10^{-5}}$. (b)~Perturbations $\theta_0(x)$ of these solitons. Sharp cusp in the plot of unstable soliton perturbation at $x_r \sim 2.5D$ is its root.}
		\label{stable_unst_example}
	\end{figure}
	
	Figure~\ref{stable_unst_example} shows the example of stable soliton (solid line), unstable soliton (dashed line) and their perturbations $\theta_0(x)$. Below we focus on stable solitons and prove that their number is infinite.
	
	\section{Multiplicity of solitons}
	\label{Multisoliton}
	There are only two static solitons in the pure sine-Gordon model: kink
	\begin{equation}
	\phi_K(x) = 4 \operatorname{arctan} \mathrm{e}^{x}
	\end{equation}
	and antikink $\phi_A(x) = - \phi_K(x)$. The most general soliton solution includes spatial shifts of these two and a choice of the left vacuum: $\varphi = \phi_K(x - x_K) + 2 \pi n$. It is impossible to combine kinks and antikinks in a static chain of solitons, since they interact with energy
	\begin{equation}
	\label{interaction_between_kinks}
	E_{int}(s_1, s_2, R) = 32 s_1 s_2 \mathrm{e}^{-R} \,.
	\end{equation}
	where $s_\alpha = +1$ for a kink, $-1$ for an antikink, and $R \gg 1$ is the distance between the solitons; see~\cite{manton} and Appendix \ref{AppendKinkAntikinkEnergy}. Indeed, widely separated kink and antikink accelerate towards each other, forming a breather, while two kinks or two antikinks repulse and go to infinity. In general, the leftmost and rightmost kinks in the solitonic chain cannot be at rest because each of them mostly interacts with the nearest neighbour.
	
	To the leading order, small external potential ($\varepsilon \ll 1$) does not change the kink and antikink profiles, but affects weak forces between them. First, consider a single kink. Substituting ${\varphi_k(x) = \phi_K(x-x_k)}$ into Eq.~(\ref{hamilt}), we obtain a periodic potential
	\begin{equation}
	\label{E_delta_general}
	E_{\delta}(x_k) \equiv H[\varphi_k] - M_k = 2 \varepsilon \sum\limits_{m \in \mathbb{Z}} \frac{1}{\cosh^2(mD - x_k)}
	\end{equation}
	which pulls the kink towards the equilibrium positions at $x_k = D(m+1/2)$; we introduced the kink mass $M_k = \int dx [(\partial_x \varphi_k)^2/2 + 1 - \cos \varphi_k] = 8$. Note that at $D \gg 1$ all terms in the sum (\ref{E_delta_general}) are exponentially suppressed except for the two largest contributions from the closest $\delta$-functions of the external potential. In particular, the potential energy of the kink centered at $0 < x_k < D$ approximately equals
	\begin{equation}
	\label{E_delta_particular}
	E_{\delta}(x_k) \approx 2 \varepsilon \left( \frac{1}{\cosh^2(x_k)} + \frac{1}{\cosh^2(D - x_k)} \right)
	\end{equation} 
	with equilibrium at $x_k = D/2$.
	
	Now, we add another kink or antikink inside the interval $lD < x'_k < (l+1)D$. The total interaction energy of the soliton pair is now\footnote{We take $l>1$, so that interaction energy of each soliton with the $\delta$-functions is not affected by another soliton.} $E_2(x_k, x'_k) = E_\delta(x_k) + E_\delta(x'_k) \pm 32 \mathrm{e}^{-(x'_k-x_k)}.$
	It is clear that if $l$ is large enough, the interaction between the (anti)kinks is exponentially small, and they remain close to the original equilibrium positions at $x_k=D/2$ and ${x'_k = (l+1/2)D}$. At small $l$ interaction between the solitons pulls them out of their potential wells, destabilizing the pair. Direct minimization of $E_{2}(x_k, x'_k)$ shows that the kink-kink and kink-antikink pairs exist at 
	\begin{equation}
	l > \frac{1}{D} \ln \frac{\varepsilon}{54} - 1 \qquad \text{and} \qquad 	l > \frac{1}{D} \ln \frac{\varepsilon}{2} + 1 ,
	\end{equation} respectively. Recall that the soliton pairs are not static in the original sine-Gordon model, so this is a new property that already can be traced back to the nonintegrability of the analogous mechanical system. Appearance of soliton pairs produces steps in the exponential growth rate of stable soliton multiplicity, see Fig.~\ref{LnNsol(n)}b. In particular, the leftmost arrow in this figure corresponds to the threshold $\varepsilon = 2 \mathrm{e}^{-D}$ for the existence of kink-antikink pair with $l = 2$.
	
	Let us demonstrate exponential growth of the stable soliton multiplicity with their length at small~$\varepsilon$. To this end consider configuration of $N$ (anti)kinks,
	\begin{equation}
	\label{sum_of_solitons}
	\varphi = \sum_{\alpha=1}^{N} s_\alpha  \varphi_k (x - x_\alpha) \,,
	\end{equation}
	where $s_\alpha = \pm 1$ distinguishes kinks from antikinks, and these objects are placed in the intervals $j_\alpha D < x_\alpha < (j_\alpha+1)D$. We will assume that distances between the adjacent (anti)kinks are large enough, so that the condition
	\begin{equation}
	\label{k_0}
	j_{\alpha+1} - j_{\alpha} > p, \qquad \text{with} \qquad p = -\frac{1}{D} \ln \frac{\varepsilon}{32} + 1
	\end{equation}
	is satisfied. In this case the interaction energy \eqref{interaction_between_kinks} between the adjacent (anti)kinks ${|E_{int}| \leqslant 32 \mathrm{e}^{-(p-1)D}}$ is at least twice smaller than their potential wells produced by the $\delta$-functions. 
	This implies that the total energy of the chain has a local minimum with respect to the position of every kink, i.e. the stable equilibrium exists. Thus, the solitons can be arbitrarily added to the chain at distances exceeding $pD$.
	{\sloppy
		
	}
	We denote the number of the above ``sparse'' solitonic chains of length $lD$ or smaller by $N_{p}(l)$, where $l$ is an integer. In the larger interval of length $(l+p)D$ one can add a kink, an antikink or none of them to the chain. Thus, 
	$
	N_{p}(l+p) \geqslant 3 N_{p}(l) .
	$
	Using $N_p(1)=3$ as the initial condition, we find\footnote{The number of sparse solitonic chains inside the interval of length $(l+p)D$ is, in fact, greater than $3 N_p$. Indeed, one can start with a chain of length smaller than $lD$ and add an (anti)kink at various positions. Taking this effect into account, one obtains more accurate recurrence relation: $N_p(l) - N_p(l-1) = 2 N_p(l-p)$ with an exponentially growing solution for $N_p(l)$. However, the latter approach also considers  ``sparse'' solitonic chains and therefore significantly underestimates the exponential growth rate $h_S$.} that 
	$
	N_p(l) \geqslant 3^{(l+p-1)/p},
	$
	i.e. the multiplicity of solitons grows at least exponentially with their length $lD$
	
	On the other hand, the total number of solitons is bounded from above by the number  $3^l$ of all possible soliton equilibrium positions with (anti)kinks occupying individual periods inside the interval $lD$. As a consequence, the number of stable solitons grows exponentially, see Eq.~\eqref{eq:2}, and the growth rate $h_S$ is bounded by
	\begin{equation}
	\label{growthratelimits}
	\frac{\ln 3}{p} \leqslant  h_S \leqslant \ln 3\,,
	\end{equation}
	where $p$ was introduced in Eq.~\eqref{k_0}.
	
\begin{figure}[h]
		\centering
		\includegraphics[scale=1]{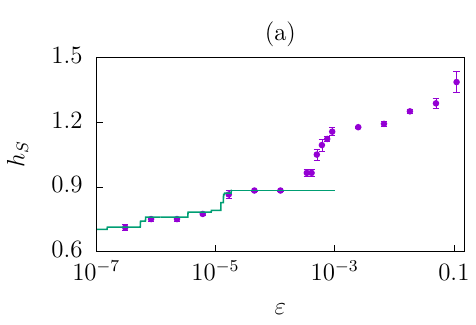}
		\includegraphics[scale=1]{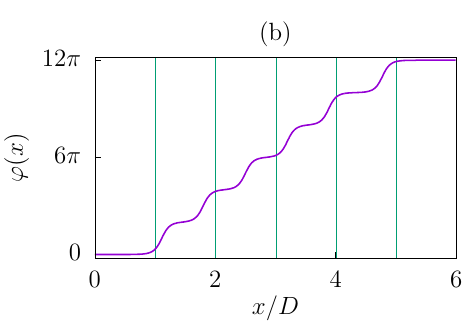}
		\vspace{-0.35cm}
		\caption{(a) Exponential growth rate $h_S(\varepsilon)$ of stable solitons. Points with errorbars are obtained by counting the number of numerically computed solitons. The solid line is found by minimizing the energy~\eqref{E_sum}. (b) An example of the soliton at $\varepsilon = 10^{-2}$ which is not represented by the ansatz (\ref{sum_of_solitons}).}
		\label{alphaeps}
	\end{figure}

	We numerically computed the number of stable solitons $N_{sol}(l)$ within the interval of length $lD$, see Fig. \ref{LnNsol(n)}a. The multiplicity indeed grows exponentially, although the growth rate $h_S(\varepsilon)$ is much higher than our lower bound~\eqref{growthratelimits}, see Figs.~\ref{LnNsol(n)}b and \ref{alphaeps}a.
	
	The next step is to consider larger $\varepsilon$ corresponding to mostly chaotic dynamics of the analogous mechanical system, see Fig.~\ref{poincare}b.
	In this case we use general expression for the energy of the solitonic chain,
	\begin{equation}
	\label{E_sum}
	E_N(x_1, \ldots, x_N) = \sum_{\alpha=1}^{N-1} E_{int}(s_\alpha, s_{\alpha+1}, x_{\alpha+1} - x_\alpha) + \varepsilon \sum_m \left(\cos \left(\varphi(mD) \right) - 1 \right) ,
	\end{equation}
	because when two (anti)kinks occupy adjacent periods of $U(x)$, they can affect each other's interaction with the external potential. 
	Minimizing (\ref{E_sum}) numerically with the conjugate gradient method, we determine whether a stable soliton chain exists for a given $\{s_\alpha, j_\alpha\}$, where $j_\alpha D < x_\alpha < (j_\alpha+1)D$. The exponential growth rate obtained from numerical minimization of energy is shown by the solid line in Fig.~\ref{alphaeps}a. It coincides with the exact graph at small $\varepsilon$, but starts to deviate from it at $\varepsilon \sim 10^{-3}$. This is due to the new types of solitons appearing in the system, with two or more (anti)kinks squeezed into one period of $U(x)$. The examples of such solitons are presented in Figs.~\ref{soliton_example}b and \ref{alphaeps}b; the ansatz (\ref{sum_of_solitons}) is not valid for them. Not surprisingly, appearance of these solitons coincides with transition to chaos in the corresponding mechanical system, cf. Figs.~\ref{poincare}b,c.
	
	\section{Topological entropy}
	\label{SecTopEntropy}
	An important quantity characterizing complexity of a dynamical system is the topological entropy~\cite{adlertopentropy}. In this section we define this quantity for the analogous mechanical system\footnote{The original topological entropy was defined in systems with compact phase space. We generalize it in a straightforward way considering a particular set of trajectories and a particular sampling of the phase space.}, then use it to constrain the soliton growth rate $h_S(\varepsilon)$.
	
	Consider the solutions starting from $\varphi \to 0$ at $x \to -\infty$.
	Let us split the field values into segments 
	\begin{equation*}
	-\pi + 2 \pi n \leqslant \varphi \leqslant \pi + 2 \pi n.
	\end{equation*}
	We characterize every solution $\varphi(x)$ in the segment of length $lD$ with the sequence of regions $(n_1, \ldots, n_l)$ it visits after every period of $U(x)$, i.e. at $x = mD+0$. One can argue that the number of different sequences $N_{seq}(l)$ grows exponentially with the length of the interval $lD$. We therefore call
	\begin{equation}
	\label{top_entropy_def}
	h_T = \lim_{l \to \infty} \, \frac{\ln N_{seq}(l)}{l}
	\end{equation} 
	the topological entropy of the analogous mechanical system.
	
	The value of $h_T$ is an indicator of chaos. The above definition gives $h_T = 0$ for ${\varepsilon = 0}$. Indeed, solutions approaching the vacuum at $x \to -\infty$ include the vacuum itself and (anti)kinks at different positions, resulting in $2l+1$ sequences of length $l$. At small nonzero $\varepsilon$ the quantity $h_T$ is positive and bounded from below by the exponential growth rate $h_S$ of the soliton multiplicity. 	 
	Indeed, every stable soliton corresponds to a unique sequence of visited vacua $(n_1, \ldots, n_l)$. Thus, $N_{sol}(l) \leqslant N_{seq}(l)$, implying~\eqref{eq:3}.
	\section{Fractal structure}
	\label{SecFractal}
	In this section we study the set of values $(\varphi(0), \varphi'(+0))$ taken by the solitonic fields at $x = +0$. We consider a small vicinity of vacuum $|\varphi(0)|, |\varphi'(+0)| \ll 1$. In this case the decomposition (\ref{general_hill_solution}) applies at $x \approx +0$, where the first and second terms vanish exponentially at negative and positive $x$, respectively. Then the complete nonlinear solution can be represented as a sum $\varphi(x) \approx \varphi_L(x) + \varphi_R(x)$ of ``left'' and ``right'' parts vanishing at $x \to +\infty$ and $x \to -\infty$. In what follows we study only the ``right'' sector of solitons, with ``left'' solutions obtained by reflection $x \to -x$.
	In particular, if $\{\varphi_\alpha(x)\}$  is the set of ``right'' solitonic field values, the entire fractal in Fig.~\ref{fractal2d}a consists of points
	\begin{equation}
	\varphi_{\alpha\beta}(0) =  \varphi_{\alpha}(0) + \varphi_{\beta}(0), \qquad
	\varphi'_{\alpha\beta}(+0) = \varphi'_{\alpha}(+0) - \varphi'_{\beta}(-0).
	\end{equation}
	Details on computing the set of ``right'' solitons $\{(\varphi_\alpha(0), \varphi'_\alpha(+0))\}$ are given in Appendix~\ref{AppendFractal}.
	
	Let us explain self-similarity of the fractal in Fig. \ref{fractal2d}. Suppose the ``right'' soliton $\varphi_S$ has parameter $A = A_S$ in Eq.~\eqref{left_condition} and length $lD$. Solution in its tiny vicinity can be represented as
	\begin{equation*}
	\varphi_A(x) = \varphi_S(x) + (A - A_S) \, \theta^{(S)}_0(x),
	\end{equation*}
	where $\theta^{(S)}_0$ is the perturbation~\eqref{Theta_0} in the background of $\varphi_S(x)$.
	Taking\footnote{We assume that $l'$ is large enough for $|\theta_0(x)|$ to reach maximum at the rightmost point $x = l'D$ of the interval.} ${A - A_S = \left[\theta^{(S)}_0 (l'D)\right]^{-1}}$ with $l' > l$, one obtains the solution $\varphi_A(x)$ staying close to $\varphi_S(x)$, arriving to the same vacuum $\varphi_n$ and then departing from it at $x > l'D$.
	At $x \approx l'D$ the solution has the form $\varphi_A(x) \approx (A - A_S) \, \theta^{(S)}_0(l'D) \, f_A(x) \, \mathrm{e}^{x - l'D} + \varphi_n$, where the asymptotics of $\theta^{(S)}_0(x)$ at $x \to +\infty$ was used. Thus, at $x = l'D$ the boundary condition \eqref{left_condition} is satisfied, with $(A-A_S) \theta^{(S)}_0(l'D)$ playing the role of the new parameter $A$. This mechanism is illustrated in Fig.~\ref{fractal2d}b and the right parts of Figs.~\ref{selfsimilarsolst}b,~c. Note that the function $\lambda_S(x) \equiv \ln \left|\theta^{(S)}_0(x)\right|$ describes exponential growth of the perturbation and therefore is related to the Lyapunov exponent of the soliton $\varphi_S(x)$.
	
	Now, let us compute the box-counting dimension of the fractal formed by the field values of solitons. To warm up, consider the entire set of solitons from the ``right'' sector, stable and unstable. This set is dense in the chaotic region, and its fractal dimension is $1$. Indeed, consider two close solutions $\varphi_1$ and $\varphi_2$ parametrized by $A_1$ and $A_2$. If the chaos is on, they diverge exponentially, with $|\varphi_1(x) - \varphi_2(x)| > 2 \pi$ at sufficiently large $x$. Then by continuity there exists a trajectory with parameter $A_S \in (A_1, A_2)$ that arrives precisely to the vacuum between $\varphi_1(x)$ and $\varphi_2(x)$. This trajectory is a soliton, which proves the statement.
	
	The parameters ${A}$ of stable ``right'' solitons, however, form a Cantor-like set with fractal dimension less than $1$. Indeed, consider the soliton $\varphi_S(x)$ with $A=A_S$. We already argued that it contains the entire set of solitons in its arbitrarily small vicinity $|A - A_S| \ll 1$, and, in particular, unstable solitons.	
	However, the solutions near the unstable solitons are also unstable: they also have roots of $\theta_0(x)$. Thus, the field values of stable soliton do not form a dense set, as their vicinities $|A-A_S| \ll 1$ contain infinitely many voids representing unstable solutions.
	
\begin{figure}[h]
		\centering
		\includegraphics[scale=1]{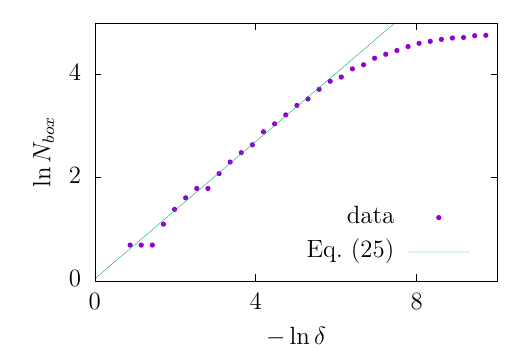}\vspace{-0.2cm}
		\caption{Number of boxes with stable solitons versus the box size at $\varepsilon = 3 \times 10^{-7}$. Fit with Eq.~\eqref{boxcountingdef} (line) gives box-counting dimension $d_{R} = 0.665 \pm 0.020$.}
		\label{howtocountboxes}
	\end{figure}

	To compute the fractal dimension we use parameter 
	\begin{equation}
	\label{a_def}
	a = \dfrac{\ln A}{\lambda_vD}.
	\end{equation} 
	instead of $A>0$.
	Since $A(a)$ is a smooth function, this does not alter fractal dimension. Transformation $A \to A \mathrm{e}^{\lambda_v D}$ trivially shifts the solution by one period of the external potential and changes ${a \to a+1}$. Thus, the fractal is periodic in $a$; in what follows we consider only the segment $a \in [0,1)$. Dividing this segment into small boxes of size $\delta$, we count the number $N_{box}(\delta)$ of boxes with stable soliton parameters $\{a_S\}$ inside.
	The details on this procedure are given in Appendix \ref{AppendFractal}.
	The box-counting fractal dimension $d_{R}$ then can be extracted from the asymptotics
	\begin{equation}
	\label{boxcountingdef}
	\ln N_{box}(\delta) \to -d_{R} \ln \delta \quad \text{as} \quad \delta \rightarrow 0.
	\end{equation} 
	The function $N_{box}(\delta)$ is shown in Fig.~\ref{howtocountboxes}.
	
	Note that the fractal dimension can be analytically bounded from below. Consider the set of stable ``right'' solitons of length $lD$ or smaller. Perturbations $\theta_0(x)$ in their backgrounds grow with $x$ at a slower rate than the vacuum perturbations, $|\theta_0(x)| \leqslant \mathrm{e}^{\lambda_v x} |f_A(x)|$, simply because $\cos \varphi(x)$ in the equation $\hat{L}_\varphi \theta_0 = 0$ is maximal at $\varphi = 2 \pi n$. As a consequence, the solitons cannot have $A$ parameters at a distance closer than
	\begin{equation}
	\label{A_size}
	\delta A_l = \left[\max \theta_0(x)\right]^{-1} \geqslant \mathrm{e}^{-\lambda_v l D}, 
	\end{equation}
	where the maximum is taken within the interval $0 \leqslant x \leqslant lD$. If Eq.~\eqref{A_size} is not satisfied, the solitons would coincide in the entire interval. This gives the typical distance between the $a$ parameters of the solitons,
	\begin{equation}
	\label{a_size}
	\delta a_l \gtrsim \mathrm{e}^{-\lambda_v l D}
	\end{equation}
	for $A \lesssim O(1)$. Breaking the $a$-interval into the boxes~\eqref{a_size}, one obtains
	\begin{equation}
	d_{R} \geqslant \lim_{l\to\infty} \frac{\ln N_{sol}(l)}{-\ln \delta a_l} \geqslant \frac{h_S}{\lambda_v D},
	\end{equation} 
	where we used Eq.~\eqref{eq:2}.
	Note that this bound is a serious underestimation: for ${\varepsilon = 3 \times 10^{-7}}$ it gives $d_{R} \gtrsim 0.06$, an order of magnitude smaller than the actual fractal dimension. Nevertheless, it proves that the dimension of our fractal is nonzero.
	
	Since the fractal in Fig.~\ref{fractal2d} is a direct sum of ``left'' and ``right'' fractals, its dimension is $d(\varepsilon) = 2d_R(\varepsilon)$, see Fig.~\ref{dbox(eps)}.
	\section{Metric entropy}
	\label{SecKolmEntropy}
	Metric (Kolmogorov-Sinai) entropy \cite{ottchaos} is an important quantity indicating whether the dynamical system is chaotic or not.
	It was originally introduced for systems with compact phase space. 
	Since our analogous mechanical system does not have this property~\cite{phasespace_noncompact}, we first modify the entropy definition as follows. 
	
	We again restrict ourselves to the ``right'' solutions of length $L = lD$ starting from $\varphi \approx 0$ at $x=0$. Besides, we consider only a finite interval $|A| \leqslant A_0$ of their shooting parameter.
	We divide the phase space into strips:
	\begin{equation}
	\label{sampling_for_Kolmogorov}
	2 \pi \nu \leqslant \varphi + \varphi' \left(\lambda_v - \frac{f'_B(+0)}{f_B(0)}\right)^{-1} < 2 \pi (\nu+1) \,,
	\end{equation}
	cf. Eq.~(\ref{right_condition}).
	For every solution $\varphi(x)$ of length $lD$ we construct the sequence $\omega = (\nu_1, \nu_2, \ldots, \nu_l)$ of visited regions at the start of every period $x = mD+0$. This divides the interval $-A_0 \leqslant A \leqslant A_0$ of solution parameters into the regions $T_\omega$ corresponding to certain sequences. The solitons belong to the boundaries of $T_\omega$ due to Eq.~\eqref{right_condition}.
	We define the metric entropy $K$ as
	\begin{equation}
	\label{kolmdef}
	K_l = -\sum_{\omega} \frac{\Delta A\left(T_\omega\right)}{2 A_0} \ln \left(\frac{\Delta A\left(T_\omega\right)}{2 A_0} \right), \qquad \text{and} \qquad 	K = \lim\limits_{l \rightarrow +\infty} \frac{K_l}{l},
	\end{equation} 
	where $\Delta A(T_\omega)$ is the total length of $T_\omega$.
	The only difference from the original Kolmogorov-Sinai construction is that we considered a selected set of trajectories and a particular sampling \eqref{sampling_for_Kolmogorov}.
	
	Note that at $\varepsilon = 0$ only two non-trivial ``right'' solutions exist, the kink and the antikink, which belong to the regions with $\nu = 0$ and $1$, respectively, at every $x$. We obtain only two sequences. Hence, $K = 0$, as it should be in the integrable case.
	
	Let us introduce the quantity analogous to the metric entropy considering the stable ``right'' solitons of length $L < lD$. Indeed, their shooting parameters divide the segment $-A_0 \leqslant A \leqslant A_0$ into multiple intervals $R_\alpha$. We therefore define
	\begin{equation}
	E_l = -\sum_\alpha \frac{\Delta A\left(R_\alpha\right)}{2 A_0} \ln \left(\frac{\Delta A\left(R_\alpha\right)}{2 A_0}\right) \qquad \text{and} \qquad E = \lim_{l\rightarrow \infty} \frac{E_l}{l},
	\end{equation}
	cf.~\eqref{kolmdef}. Clearly, $E$ characterizes (in)homogeneity of distribution of the stable soliton shooting parameters. If all solitons have the same $A$, then $E = 0$. If they are evenly distributed, then $E_l = \ln N_{sol}(l)$ and $E = h_S$, see Eq.~\eqref{eq:2}.
	
\begin{figure}[h]
		\centering
		\includegraphics[scale=1]{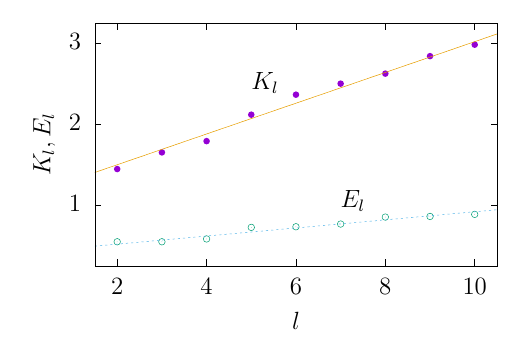}
		\vspace{-0.3cm}
		\caption[footnote]{Values of $E_l$ and $K_l$ computed at $\varepsilon = 3 \times 10^{-7}$ and $A_0 \approx 2.5 \times 10^{-6}$. Linear fit gives $K = 0.19 \pm 0.01$ and $E = 0.050 \pm 0.004$.}
		\label{KolmForSmallEps}
	\end{figure}

	Since the boundaries of $R_\alpha$ are also the boundaries of $T_\omega$, splitting $\{R_\alpha\}$ is a coarse-graining of $\{T_\omega\}$ obtained by merging some of the regions together. However, if two regions of lengths $\Delta A_1$ and $\Delta A_2$ are merged into one,
	\begin{equation*}
	-(\Delta A_1 + \Delta A_2) \ln\left(\frac{\Delta A_1 + \Delta A_2}{2A_0}\right) \leqslant -\Delta A_1 \ln	\left(\frac{\Delta A_1}{2A_0}\right) -\Delta A_2 \ln \left(\frac{\Delta A_2}{2A_0}\right).
	\end{equation*}
	This proves that $E_l \leqslant K_l$, and therefore $E \leqslant K$. In Fig.~\ref{KolmForSmallEps} we demonstrate the values of $E_l$, $K_l$ (points) and their linear fits (lines). 
	
	Thus, one can use stable solitons in field theory to constrain metric entropy of the analogous mechanical system.
	
	\section{Generalization}
	\label{SecDiscuss}
	In this paper we studied solitons in one-dimensional theories with nonintegrable static field equations. Specifically, we considered the sine-Gordon model in Dirac comb potential. This choice allowed us to implement an efficient numerical method for computing the multisoliton solutions. Due to the chaotic nature of the equations there exists an infinite set of these objects. Besides, we have found that the field values of the solitons form a fractal in the configuration space. We computed non-integer box-counting dimension of the fractal and studied it using the metric and topological entropies.
	
	We do not want to leave an impression, however, that our model is special in some regard. Similar ``chaotic'' solitons should exist in many one-dimensional theories with non-integrable static equations, cf.~\cite{yamashitachain, nonlinschrod}. The simplest generalizations include one-field models with different periodically driven potentials $V(\varphi, x)$. If several $x$-independent degenerate vacua are present, these models possess topological solitons interpolating between the vacua. The soliton-counting method of this paper is then applicable if the periodic driving of the potentials is sufficiently weak and its period $D$ exceeds the width of the elementary ``kink-like'' solitons. In this case the number of solitons should grow exponentially with their length, and the solitonic field values should form self-similar fractals in the configuration spaces, just like in our model.
	
	Another set of one-field models involves potentials $V(\varphi, x)$ with non-periodic spatial dependence. For example, one can consider the same driven sine-Gordon model, but with $\delta$-functions placed non-periodically in Eq.~\eqref{perturb_as_delta}, at $x = x_m \neq mD$. If $\varepsilon$ is moderately small and the distance $x_{m+1} - x_m$ between the neighboring $\delta$-functions exceeds the kink width, the solitonic chains in this model can be constructed in the same way as in Sec.~\ref{Multisoliton}. Then there should exist an infinite number of stable solitons. Besides, their multiplicity should grow exponentially with the number $l$ of $\delta$-functions inside the soliton profile: expression \eqref{eq:2} with $l$ in place of $L/D$. The field values of these solitons should form complicated hierarchical structures. However, self-similarity observed in Fig.~\ref{fractal2d} should be broken. Indeed, the argument of Sec.~\ref{SecFractal} relates magnifications of the soliton vicinity in the $\varphi(0), \varphi'(0)$ plane to spatial translations of the soliton parts. If the discrete translation symmetry is broken, the self-similarity should disappear. Nevertheless, the box-counting fractal dimension of the set $\{\varphi(0), \varphi'(0)\}$ may be non-integer.
	
	An interesting special case is obtained by placing $\delta$-functions in Eq.~\eqref{perturb_as_delta} at random positions $x = x_m$. This may represent some kind of impurities in the original sample. Every realization of $\{x_m\}$ in this case corresponds to non-periodic $\delta$--comb. However, averaging over the random ensemble may essentially change the final properties of the solitons. The study of this notable case is beyond the scope of the present paper.
	
	Let us point out that the results of this paper can be extended at least to some class of multifield models. Indeed, consider two fields $\varphi_1$ and $\varphi_2$ with the energy functional
	\begin{equation}
	H = \Gamma H_1 [\varphi_1] + H_2 [\varphi_1, \varphi_2]\,,
	\end{equation}
	where $\Gamma$ is a constant. Equations for the static solitons are
	\begin{equation}
	\label{dH/dphi12}
	\frac{\delta H_1[\varphi_1]}{\delta \varphi_1} +\frac{1}{\Gamma}\frac{\delta H_2 [\varphi_1, \varphi_2]}{\delta \varphi_1} = 0\,, \qquad \frac{\delta H_2 [\varphi_1, \varphi_2]}{\delta \varphi_2} = 0\,.
	\end{equation}
	At $\Gamma \gg 1$ the field $\varphi_1$ satisfies an independent equation, while $\varphi_2(x)$ evolves in the external potential $\varphi_1(x)$. If the latter is periodic, the properties of $\varphi_2$--solitons may be close to those in our model.
	
	Generically, one expects to find an infinite total number of solitons in non-integrable case. But the distribution of the stable solitons may be model-dependent. 
	Indeed, these solutions are the local minima of the static energy $H[\varphi]$ which coincides with the classical action in the mechanical analogy. Presently, there is no general classification of mechanical trajectories locally minimizing the action, though some works in this direction appear \cite{leastornot}.
	
	We hope that the instruments developed in this paper --- metric and	topological entropies, and fractals formed by field values ---
	will be useful for studies of chaotic solitons in different theories.
	\paragraph*{Acknowledgements.}
	We are grateful to V.~A.~Rubakov for comments and criticism. This work was supported by the grant RSF 16-12-10494. Numerical calculations were performed on the Computational Cluster of the Theoretical Division of INR RAS.
	
	\appendix
	\section{Deriving asymptotic conditions}
	\label{appendix_asymptotic}
	Near the vacuum $\varphi = 0$ Eq.~(\ref{osc+perturb}) takes the form of a Schr\"odinger equation	
	\begin{equation}
	\label{append_linearized}
	\varphi''(x) = \varphi(x) U(x)
	\end{equation}
	in the periodic potential $U(x)$, Eq.~(\ref{perturb_as_delta}). The particular solutions of this equation coincide with the eigenfunctions of the shift operator $x \rightarrow x + D$. This suggests the ansatz
	\begin{equation}
	\varphi(x) = \mathrm{e}^{\lambda x} f(x)\,,
	\end{equation}
	where $f(x)$ has period $D$.
	Solving Eq.~\eqref{append_linearized} inside the interval $0 < x < D$, one obtains,
	\begin{equation}
	\label{gen_sol_f}
	f(x) = \mathrm{e}^{-\lambda x} (C_+ \mathrm{e}^x + C_- \mathrm{e}^{-x})\,.
	\end{equation}
	The linearized matching conditions \eqref{matching_sol} at the endpoints of this interval take the form
	\begin{equation}
	\label{asymp_f}
	f(0) = f(D), \qquad f'(D) + \varepsilon f(0) = f'(0) \,,
	\end{equation}
	where we recalled that $f$ is periodic.
	Substituting Eq.~(\ref{gen_sol_f}) into Eq.~\eqref{asymp_f}, we arrive to the homogeneous linear system
	\begin{equation}
	\label{matrix_f}
	\left( 
	\begin{matrix}
	\mathrm{e}^{(1-\lambda)D}-1 & \mathrm{e}^{-(1+\lambda)D}-1 \\\varepsilon+\lambda -1+ (1-\lambda) \mathrm{e}^{(1-\lambda)D} & \varepsilon+\lambda +1-(1+\lambda)\mathrm{e}^{-(1+\lambda)D}
	\end{matrix}\right)\left( 
	\begin{matrix}
	C_+ \\ C_-
	\end{matrix}
	\right) = 
	0\,,
	\end{equation}
	which has nontrivial solutions only if the matrix has zero determinant. This gives two roots
	\begin{equation}
	\label{lambda_v}
	\lambda = \pm \frac{1}{D} \ln \left( \sigma + \sqrt{\sigma^2-4}\right) \equiv \pm \lambda_v\,, \qquad \sigma = \cosh D + \frac{\varepsilon}{2} \sinh D \,.
	\end{equation}
	We obtained Eq.~(\ref{general_hill_solution}), where the particular solutions $f_A$, $f_B$ are given by Eq.~(\ref{gen_sol_f}) with $\lambda = \pm \lambda_v$ and coefficients $C_+$, $C_-$ representing the eigenvectors in Eq.~\eqref{matrix_f}. We normalize the solutions by $C_+ + C_- = 1$.
	\section{General solution of the static sine-Gordon equation}
	\label{appendix_solving}
	In the regions between the $\delta$-functions Eqs.~(\ref{osc+perturb}), (\ref{perturb_as_delta}) reduce to the equation for physical pendulum: $\varphi'' = \sin \varphi$.
	The motion of the latter system depends on whether its mechanical energy 
	\begin{equation}
	\label{mech_energy}
	\mathscr{E}_{m} = \frac{\varphi'^2}{2} + \cos \varphi - 1 
	\end{equation}
	exceeds the height of the barrier $\mathscr{E}_{m} = 0$ or not \cite{zaslavskyoldbook}. Periodic motions at $\mathscr{E}_{m} < 0$ are described by general solution
	\begin{equation}
	\label{phi_osc}
	\varphi(x) = 2 \operatorname{arccos} (\pm k\, \operatorname{sn} (x-x_{\pi}, k)) + 2 \pi n\, , \quad n\in \mathbb{Z}\,,
	\end{equation}
	while the ``rotating'' solutions at $\mathscr{E}_{m} > 0$ are
	\begin{equation}
	\label{phi_rot}
	\varphi(x) = \pi \pm 2 \operatorname{am} \left(k(x-x_\pi), \, \frac{1}{k}\right)\,.
	\end{equation}
	Here $\operatorname{sn}$ and $\operatorname{am}$ denote the elliptic sine and Jacobi amplitude, respectively \cite{nist}. The solutions \eqref{phi_osc}, \eqref{phi_rot} have two integration constants: $k = \sqrt{\mathscr{E}_{m}/2 + 1}$ and the shift parameter $x_\pi$. Signs $\pm$ in these equations discriminate two branches of solutions with opposite signs of $\varphi'(x)$.
	
	Numerically, we use Eqs.~\eqref{phi_osc}, \eqref{phi_rot} as follows. Starting with the values of $\varphi$ and $\varphi'$ at $x = mD+0$, we compute $\mathscr{E}_m$, $k$, and determine the relevant branch of the general solution. Inverting Eq.~\eqref{phi_osc} or \eqref{phi_rot}, we find $x_\pi$ and hence --- values of $\varphi$ and $\varphi'$ within the entire interval ${mD < x < (m+1)D}$. Using the matching conditions (\ref{matching_sol}), we proceed with the next interval.
	\section{Linear stability}
	\label{appendix_stability}
	Let us describe a practical way to study soliton stability within the shooting approach. To this end we count zeros of the perturbation $\theta_0(x)$ in Eq.~(\ref{Theta_0}).
	
	As the shooting parameter $A$ changes, these zeros cannot disappear or emerge sporadically inside the interval $0 < x < L$. Indeed, one can regularize the $\delta$-functions in Eq.~\eqref{perturb_as_delta}, making $\theta_0$ a smooth function of $x$ and $A$. After that the roots of $\theta_0$ can appear at the real axis or disappear from it only in pairs at points $x_*$ such that $\theta_0(x_*) = \theta'_0(x_*) = 0$. However, $\theta_0(x)$ is non-trivial and satisfies the second-order linear equation $\hat{L}_\varphi \theta_0 = 0$. It cannot vanish together with its first derivative at any $x$.
	
	As a consequence, the number of $\theta_0$ roots inside the interval $0<x<L$ changes by~$\pm 1$ when one of them crosses $L$. This happens at certain values $A = A_*$ of the shooting parameter satisfying $\partial \varphi(L)/\partial A = 0$. Thus, we just need to find out whether the root $x_r(A)$ comes in or goes out of the interval. Differentiating the equality $\theta_0(x_r(A), A)=0$, we obtain the change $\Delta N_r$ in the number of roots,
	\begin{equation}
	\label{where_the_root_goes}
	\Delta N_r = - \operatorname{sgn}\left.\frac{dx_{r}}{dA}\right|_{A_*} = \operatorname{sgn} \left.\dfrac{\partial_{A} \theta_{0}(L)}{\theta'_{0}(L)}\right|_{A_*}\,.
	\end{equation}
	Expression~(\ref{where_the_root_goes}) can be simplified using the mechanical energy $\mathscr{E}_m$, Eq.~(\ref{mech_energy}), which does not depend on $x$ inside the finite intervals $mD < x < (m+1)D$. Using Eq.~(\ref{Theta_0}), one finds
	$
	\partial_x \theta_{0} (L)  = \partial_x \partial_{A} \varphi(L)  = \partial_A \mathscr{E}_{m}/\varphi'(L)
	$
	at $A = A_*$. Therefore,
	\begin{equation}
	\label{root_path_from_H}
	\Delta N_r =  \left. \operatorname{sgn} \left(\partial_{A} \mathscr{E}_{m} \cdot \partial^2_{A} \varphi(L) \cdot \varphi'(L)\right)\right|_{A_*} \,.
	\end{equation}
	The factors in Eq.~(\ref{root_path_from_H}) can be computed numerically using the values of $\varphi$ and $\varphi'$ at $x = L$ and different $A$. The latter are provided by the shooting method.
	
	We apply Eq.~(\ref{root_path_from_H}) as follows. At small $A$ the solutions $\varphi_A(x)$ remain close to the vacuum, and $\theta_0(x)$ does not have zeros, $N_r = 0$. We change $A$ in small steps and determine the values $A_*$ corresponding to $\partial_A \varphi(L)=0$. At these points we change the number of $\theta_0$ roots according to Eq.~(\ref{root_path_from_H}). Once all solitons are obtained, we select the stable ones, i.e. those with $N_r = 0$.
	
	We tested the above procedure by explicitly solving the equation $\hat{L}_\varphi \theta_0 = 0$ via the sequential algorithm, cf. Appendix~\ref{appendix_solving}, arriving to the same result for the number of $\theta_0$ zeros.
	
	\section{Interaction energy of a soliton pair}
	\label{AppendKinkAntikinkEnergy}
	Consider a kink and an (anti)kink in the pure sine-Gordon model with centers separated by distance $R \gg 1$. This field configuration is approximated by a sum
	\begin{equation}
	\label{kink_(anti)kink}
	\varphi_{2}(x) \equiv 4 \arctan \mathrm{e}^{x + R/2} \pm  4 \arctan \mathrm{e}^{x - R/2} = \varphi_l(x) \pm \varphi_r(x)\,,
	\end{equation}
	where plus and minus signs correspond to kink and antikink at $x = R/2$, respectively.
	Substituting (\ref{kink_(anti)kink}) into the energy \eqref{hamilt} at $\varepsilon = 0$, we find,
	\begin{equation}
	H = -16 \int\limits_{-\infty}^{+\infty} \frac{dx (1 \mp \cosh 2x)}{(\cosh 2x + \cosh R)^2} + \operatorname{const} = \pm 32 \mathrm{e}^{-R} + O(\mathrm{e}^{-2R}) + \operatorname{const},
	\end{equation}
	where the constant includes all $R$-independent terms. In the last equality we computed the integrals and extracted the asymptotics $R \to \infty$, reproducing the well-known result~\cite{manton}.
	
	\section{Finding the fractal to a given precision}
	\label{AppendFractal}
	In a nutshell, our numerical procedure for computing the fractal of ``right'' solitonic values (Sec.~\ref{SecFractal}) is straightforward. We split the values of $a \in [0, 1]$ into small boxes of size $\delta$, search for stable solitons within each box by the shooting method, and plot their field values with points in Figs.~\ref{fractal2d}a,b. Changing the box size $\delta$, we obtain Fig.~\ref{howtocountboxes} and the coefficient in Eq.~\eqref{boxcountingdef}.
	
	To calculate the fractal with resolution $\delta$, however, we need to search for the solitons of different length $L=lD$ in different $a$-boxes. We estimate $L$ by recalling that the difference between any two solutions grows exponentially with $x$:
	$\Delta \varphi \sim \Delta A \cdot \theta_0(x) \,.	$
	Thus, for a given soliton length $L$ we take the interval $\Delta A = \left[\theta_{\max}(L)\right]^{-1}$, where $\theta_{\max}(L)$ is the maximum of $|\theta_0(x)|$ at $0<x<L$. Solutions within this interval satisfy $\Delta \varphi \lesssim 1$, so it contains $O(1)$ stable solitons of length $L$, if they exist. Inversely, for a given $\Delta a = \delta$ we take large enough $L$ satisfying
	\begin{equation}
	\label{theta_condition}
	\theta_{\max}(L) \geqslant \frac{1}{\Delta A} \geqslant \frac{\mathrm{e}^{-\lambda_v D \cdot a_0}}{\delta \cdot \lambda_v D} \,,
	\end{equation}
	where in the last expression we converted $\Delta A$ into $\Delta a = \delta$ and used Eq.~\eqref{a_def}.
	
	In practice we compute $\theta_0(x)$ for the solution in the center of each $a$-box to the point $x=L$ where Eq.~\eqref{theta_condition} is already satisfied, then search for solitons of length $L$ within this box.

\end{document}